\documentclass[english,aps,pra,superscriptaddress,twocolumn]{revtex4-1}
\usepackage{mathptmx}
\usepackage[T1]{fontenc}
\usepackage[latin9]{inputenc}
\usepackage{color}
\usepackage{babel}

\usepackage{verbatim}
\usepackage{amsmath}
\usepackage{graphicx}
\usepackage{amssymb}
\usepackage{esint}
\usepackage[unicode=true,pdfusetitle,
 bookmarks=true,bookmarksnumbered=false,bookmarksopen=false,
 breaklinks=false,pdfborder={0 0 1},backref=false,colorlinks=false]
 {hyperref}
\usepackage{breakurl}

\makeatletter
\@ifundefined{textcolor}{}
{%
 \definecolor{BLACK}{gray}{0}
 \definecolor{WHITE}{gray}{1}
 \definecolor{RED}{rgb}{1,0,0}
 \definecolor{GREEN}{rgb}{0,1,0}
 \definecolor{BLUE}{rgb}{0,0,1}
 \definecolor{CYAN}{cmyk}{1,0,0,0}
 \definecolor{MAGENTA}{cmyk}{0,1,0,0}
 \definecolor{YELLOW}{cmyk}{0,0,1,0}
 }

\makeatother

\begin{document}

\title{Giant Photogalvanic Effect in Metamaterials \\
Containing Non-Centrosymmetric Plasmonic Nanoparticles }

\author{Sergei V. Zhukovsky}

\email{sezh@fotonik.dtu.dk}

\affiliation{DTU Fotonik -- Department of Photonics Engineering, Technical University
of Denmark, {\O}rsteds Plads 343, DK-2800 Kgs. Lyngby, Denmark}

\affiliation{National Research University of Information Technology, Mechanics
and Optics, Kronverksky pr. 49, St. Petersburg, 197101, Russia}

\author{Viktoriia E. Babicheva}

\affiliation{National Research University of Information Technology, Mechanics
and Optics, Kronverksky pr. 49, St. Petersburg, 197101, Russia}

\affiliation{DTU Fotonik -- Department of Photonics Engineering, Technical University
of Denmark, {\O}rsteds Plads 343, DK-2800 Kgs. Lyngby, Denmark}

\affiliation{Birck Nanotechnology Center, Purdue University, 1205 West State Street,
West Lafayette, IN, 47907-2057 USA}

\author{Andrey B. Evlyukhin}

\affiliation{Laser Zentrum Hannover e.V., Hollerithallee 8, D-30419 Hannover,
Germany}

\author{Igor E. Protsenko}

\affiliation{P. N. Lebedev Physical Institute, Russian Academy of Sciences, Leninskiy
Pr. 53, 119333 Moscow, Russia }

\affiliation{Advanced Energy Technologies Ltd, Skolkovo, Novaya Ul. 100, 143025
Moscow Region, Russia }

\author{Andrei V. Lavrinenko}

\affiliation{DTU Fotonik -- Department of Photonics Engineering, Technical University
of Denmark, {\O}rsteds Plads 343, DK-2800 Kgs. Lyngby, Denmark}

\author{Alexander V. Uskov}

\affiliation{P. N. Lebedev Physical Institute, Russian Academy of Sciences, Leninskiy
Pr. 53, 119333 Moscow, Russia }

\affiliation{Advanced Energy Technologies Ltd, Skolkovo, Novaya Ul. 100, 143025
Moscow Region, Russia }
\begin{abstract}
Photoelectric properties of metamaterials containing non-centrosymmetric,
similarly oriented metallic nanoparticles embedded in a homogeneous
semiconductor matrix are theoretically studied. Due to the asymmetric
shape of the nanoparticle boundary, photoelectron emission acquires
a preferred direction, resulting in a photocurrent flow in that direction
when nanoparticles are uniformly illuminated by a homogeneous plane
wave. This effect is a direct analogy of the photogalvanic (or bulk
photovoltaic) effect known to exist in media with non-centrosymmetric
crystal structure, such as doped lithium niobate or bismuth ferrite,
but is several orders of magnitude stronger. Termed the \emph{giant
plasmonic photogalvanic effect}, the reported phenomenon is valuable
for characterizing photoemission and photoconductive properties of
plasmonic nanostructures, and can find many uses for photodetection
and photovoltaic applications.
\end{abstract}


\maketitle

\section{Introduction}

The recent decade in modern physics has featured the concept of optical
metamaterials. The central idea of this concept is to bestow the role
of known, ordinary constituents of matter (atoms, ions, or molecules)
upon artificial {}``meta-atoms''---nanosized objects purposely designed
to have the desired optical properties \cite{mmBook}. The assembly
of such meta-atoms---an artificial composite metamaterial---exhibits
the desired properties macroscopically, provided that the meta-atoms
are much smaller than the wavelength of light interacting with them. 

Great as the variety of naturally occurring atoms and molecules (and,
in turn, of natural materials) may be, the inherent total freedom
in choosing the shape and composition of artificial meta-atoms is
believed to be even greater---nearly arbitrary. Thus, a prominent
success of optical metamaterials is the design of materials with optical
properties that either do not exist or are much weaker in naturally
occurring media. Notable examples include metamaterials with a negative
refractive index, near-zero, or very large permittivity; metamaterials
with magnetic permeability at optical frequencies; extremely anisotropic
hyperbolic metamaterials that behave like metals in some directions
and like dielectrics in others; chiral metamaterials with giant magnetooptical
properties, and many others \cite{mm1,mm2,mm3}. 

Most meta-atom designs proposed to date are based on metallic nanoparticles,
nanoantennas, or resonators of various shapes \cite{mmAnt}. In such
metallic structures, the size prerequisite for meta-atom design is
fulfilled by subwavelength confinement of electromagnetic field due
to localized surface plasmon resonance excitation. At the same time,
localized plasmons are known to cause strong local field enhancement,
which can enhance the functionality of metamaterials in the context
of biological and chemical sensing, as well as give rise to new concepts
of optical metamaterials based on strongly enhanced nonlinear, photorefractive,
and photoconductive effects. In particular, plasmonics-enhanced photoconductivity--the
emission of photoelectrons from nanoparticles due to action of strong
local fields in the localized plasmonic resonances--was recently shown
to be promising for photodetection and photovoltaic applications \cite{NordlanderNatC13,NordlanderNL13,ourPlas}. 

Transcending the purely electromagnetic approach traditionally adopted
in the study of plasmonic nanostructures and accounting for processes
when light can cause electrons to leave the nanoparticles has far-reaching
implications, putting forth a new concept of photoconductive metamaterials.
The enhanced photoelectric effect from plasmonic nanoantennas with
generation of {}``hot'' electrons \cite{uGovorov,uReview1} can
be used to improve the characteristics of light-harvesting devices
(e.g., photoconductive plasmonic metamaterials, photodetectors, solar
and photochemical cells) \cite{uCarius13,uCatchpole12,uHalasSci11,uMisawa10,uMoskovits13,uTatsuma11,plas5-UskovUFN12,plas8-Novitsky12,uGovorov,uReview1,uAtwater14},
as well as more generally in optoelectronics, photochemistry, and
photo electrochemistry \cite{ch1-Small,ch2-Nanoscale,ch3-SREP}. 

In this paper\textbf{, }we predict and numerically demonstrate an
effect related to new functionality of photoconductive metamaterials:
the \emph{giant plasmonic photogalvanic effect}. Named after photogalvanic
(or bulk photovoltaic) effect in bulk non-centrosymmetric media \cite{bookSturmanFridkin},
plasmonic photogalvanic effect is shown to exist in a metamaterial
containing similarly oriented non-centrosymmetric metallic nanoparticles
embedded in a homogeneous semiconductor matrix. The low degree of
symmetry of the nanoparticle shape causes the net flux of the {}``hot''
electrons emitted from the nanoparticles via the resonant plasmonic
excitation to be directional. This directionality leads to a photo-electromotive
force as a result of homogeneous external light illumination (the
photogalvanic effect). 

We report that the resulting photocurrent density generated in a layer
of nanoparticles emerges and grows as the particle shape changes from
cylindrical to conical, i.e., with the increase of the particle asymmetry.
We calculate the components of the effective third-rank tensor relating
the current density to the incident electric field, and show that
this effective tensor for the nanoparticle array exceeds that for
the naturally occurring ferroelectrics that exhibit bulk photovoltaic
effect. Hence, the reported plasmonic effect can be regarded as a
{}``giant'' version of the conventional (non-plasmonic) photogalvanic
effect occurring in natural materials.

%
\begin{comment}
The reported plasmonic photogalvanic effect bridges the gap between
the inner photoelectric effect, which results from light-matter interaction
with a microscopic object such as an atom or an impurity center, and
the outer photoelectric effect, which occurs at material interfaces
in a macroscopic setting \cite{bookInnerOuter}. It is therefore fundamentally
useful in providing clues about the nature of plasmon-assisted electron
photoemission and photoinduced processes, and can have a variety of
applications in photodetection, photovoltaics, and photochemistry.%
\end{comment}
{}

\begin{figure}
\includegraphics[width=1\columnwidth]{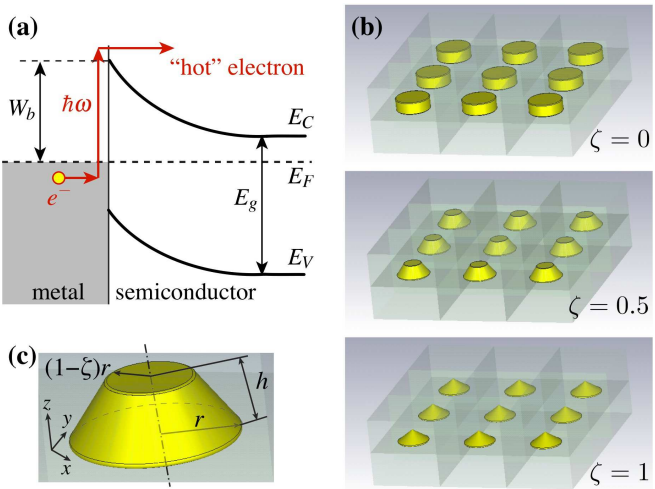}

\caption{Schematics of (a) {}``hot'' photoelectron emission through the Schottky
barrier at a metal/semiconductor interface; (b) several nanoparticle
arrays studied in the present paper (showing three characteristic
cases of cylinders, truncated cones, and cones); (c) enlarged view
of one nanoparticle showing its geometrical parameters. \label{FIG:Schematics}}
\end{figure}

The paper is organized\textbf{ }as follows.\textbf{ }In Section~\ref{sec:THEORETICAL}
we review the theoretical background for {}``hot'' electron photoemission
at metal-semiconductor interfaces containing Schottky barriers. In
Section~\ref{sec:NUMERICAL} we numerically investigate arrays of
nanoparticles whose shape varies from cylindrical to conical. We discuss
the observed increase of photocurrent directionality and induced electromotive
force as a result of increased spatial asymmetry of the nanoparticles.
In Section~\ref{sec:DISCUSSION} we draw parallels between the predicted
plasmonic photogalvanic effect and the known photogalvanic effect
in certain naturally occurring media. Finally, in Section \ref{sec:CONCLUSIONS}
we summarize and outline possible applications for the proposed effect.

\section{Theoretical background\label{sec:THEORETICAL}}

We consider a metallic nanoparticle embedded in a uniform semiconductor
matrix (Fig.~\ref{FIG:Schematics}) in presence of a normally incident
light wave of frequency $\omega$, such that \begin{equation}
W_{b}<\hbar\omega<E_{g}.\label{eq:energy}\end{equation}
It is assumed that the photon energy $\hbar\omega$ is insufficient
to excite electron-hole pairs in the semiconductor matrix with gap
energy $E_{g}$, but exceeds the work function for the metal/semiconductor
interface $W_{b}$ (see Fig.~\ref{FIG:Schematics}a), so that the
photon energy transferred to an electron in metal can cause it to
leave the nanoparticle. 

Two mechanisms of energy transfer from the photon to the electron
can be identified \cite{plas22-Tamm31,plas23-BrodskyBook,plas24-BrodskyJETP}.
One is absorption of a photon by an electron in the bulk of the nanoparticle
with subsequent transport of the \textquotedblleft{}hot\textquotedblright{}
electron to the surface and its emission by overcoming the Schottky
barrier (the volume photoelectric effect \cite{plas25-Berini10}).
The other is absorption of a photon by an electron as it collides
with the nanoparticle boundary, causing emission of that particular
electron from the metal (the surface photoelectric effect). In both
of these mechanisms, the spatial and momentum distribution of the
emitted electrons is strongly influenced by (i) the spatial configuration
of the nanoparticle surface and (ii) resonant field enhancement near
that surface due to the plasmonic resonance. 

While the question on which of the two mechanisms is stronger in nanostructures
is yet to be answered, it was shown that electron collisions after
photon absorption leading to rapid {}``hot'' electron cool-down
make surface-driven effects prevail over bulk effects in many cases
\cite{plas22-Tamm31,ourTammBulk}. So, following the previously developed
Tamm theory of photoemission from plasmonic nanoparticles \cite{plas5-UskovUFN12,plas8-Novitsky12,ourPlas},
we can express the photocurrent from a nanoparticle as \begin{equation}
I{}_{NP}(\lambda)=C_{em}(\lambda)\oint_{\text{particle}}\left|E_{n}(\mathbf{r})\right|^{2}d^{2}\mathbf{r}.\label{eq:photocurrent_scalar}\end{equation}
%
\begin{comment}
Here, the proportionality coefficient $C_{em}(\lambda)$ depends on
the properties of the Schottky barrier between metal and semiconductor,
and in particular, on the work function $W_{b}$ \cite{plas5-UskovUFN12,plas25-Berini10},
as seen in Fig.~\ref{FIG:Schematics}a. %
\end{comment}
{}The integration is performed over the surface of the nanoparticle.
The expression in Eq.~\eqref{eq:photocurrent_scalar} only takes
into account the increased number of photoelectrons without account
for the direction of their momentum as they leave the nanoparticle.
As a quantity integrated for all electrons in all directions, it is
applicable in the photoconductivity scenario when the emitted electrons
are subsequently directed using an externally applied or a built-in
potential, and their momentum direction upon leaving the nanoparticle
can thus be totally disregarded. 

In the absence of such potential, the initial velocity of the emitted
photoelectrons starts to play a role in the definition of the photocurrent
from a nanoparticle, which then has to assume a modified form, \begin{equation}
\mathbf{I}{}_{NP}(\lambda)=-C_{em}(\lambda)\oint_{\text{particle}}\left|E_{n}(\mathbf{r})\right|^{2}\mathbf{n}d^{2}\mathbf{r},\label{eq:photocurrent_vector}\end{equation}
where $\mathbf{n}$ is the unit normal vector  at the nanoparticle
surface at point $\mathbf{r}$. In both Eqs.~\eqref{eq:photocurrent_scalar}
and \eqref{eq:photocurrent_vector}, the coefficient $C_{em}$ equals
\begin{equation}
C_{em}=\eta_{o}\frac{e}{\hbar\omega}Y_{m}=\eta_{o}\frac{e}{\hbar\omega}\cdot\frac{\epsilon_{0}cn_{m}}{2},\label{eq:C_em}\end{equation}
where $n_{m}$ is the refractive index of the matrix and $\eta_{o}$
is the external quantum efficiency of the electron photoemission through
the potential barrier at the metal/matrix interface \cite{ourPlas,ourSpecIssue};
it strongly depends on the photon energy and varies from zero at $\hbar\omega=W_{b}$
(see Fig.~\ref{FIG:Schematics}a) to non-zero values for higher photon
energies \cite{ourTammBulk}. The admittance (inverse impedance) $Y_{m}$
of the matrix medium relates the intensity $S$ of a plane wave with
the electric field strength $E$ as $S=Y_{m}|E|^{2}$. 

When a plasmonic resonance is excited in the nanoparticle, the amplitude
of the resonant fields inside it and near its surface usually greatly
exceeds the amplitude of the incident field. Therefore, we can see
from Eqs.~\eqref{eq:photocurrent_scalar}--\eqref{eq:photocurrent_vector}
that in nanoparticles with centrosymmetric shapes (such as nanospheres,
nanocubes, or nanodisks) the spatial symmetry of the resonant modes
will lead to the cancellation of the directed photocurrent, so that
$\mathbf{I}_{NP}=0$ even though $I_{NP}\neq0$. In other words, even
though the incident wave can cause additional photoelectrons to be
emitted over the Schottky barrier, the collective motion of these
electrons will not induce an electromotive force.

The situation changes drastically when the shape of the nanoparticle
and/or the field distribution of the resonant plasmonic mode lacks
the center of symmetry. Eq.~\eqref{eq:photocurrent_vector} then
shows that the resulting photocurrent acquires directionality, so
that photoemission from such a non-centrosymmetric nanoparticle results
in net photocurrent ($\mathbf{I}_{NP}\ne0$) and induces electromotive
force. The ratio $\rho=|\mathbf{I}_{NP}|/I_{NP}$, which can vary
from 0 to 1, can be regarded as a measure of directionality for photoemission
with respect to one nanoparticle.

In a metamaterial comprising an oriented arrangement of such non-centrosymmetric
nanoparticles that is not too dense, the satisfactory approximation
is that the neighboring particles do not modify the plasmonic resonance
of each other in a significant way. In this case, the individual photocurrents
from each particle sum up, with the resulting current density from
a square nanoparticle lattice with period $a$ written as $\mathbf{j}=\mathbf{I}_{NP}/a^{2}$.
Since the nanoparticles are axially symmetric with respect to the
$z$-axis (see Fig.~\ref{FIG:Schematics}c), and the lattice has
4-fold rotational symmetry, we expect that for a normally incident
wave $j_{x}=j_{y}=0$ and $j_{z}=|\mathbf{j}|$, so we can write \begin{equation}
j_{z}=|\mathbf{I}_{NP}|/a^{2}=\rho I_{NP}/a^{2}=\rho C_{em}|E_{0}^{2}|\xi\label{eq:photocurrent_volume}\end{equation}
where $E_{0}$ is the field \emph{incident} on the lattice, and \begin{equation}
\xi=\frac{1}{|E_{0}^{2}|a^{2}}\oint_{\text{particle}}\left|E_{n}(\mathbf{r})\right|^{2}d^{2}\mathbf{r}\label{eq:xi}\end{equation}
relates the incident field $E_{0}$ with the local field $E(\mathbf{r})$
and has the meaning of a field enhancement factor due to the localized
plasmon resonance in the nanoparticles \cite{ourPlas,ourSpecIssue}.
If the incident field is, e.g., $x$-polarized, Eq.~\eqref{eq:photocurrent_volume}
can be rewritten as\begin{equation}
j_{z}=\rho\xi C_{em}E_{0x}E_{0x}^{*}=\tilde{\alpha}_{zxx}E_{0x}E_{0x}^{*},\label{eq:photogalvanic_tilde}\end{equation}
which has the form equivalent to the photocurrent induced in some
media with non-centrosymmetric crystal structure due to the photogalvanic
(or bulk photovoltaic) effect \cite{Fridkin2001,bulkYoungPRL,galvPhysE},\begin{equation}
j_{i}=\alpha_{ijk}E_{j}E_{k}^{*},\label{eq:photogalvanic}\end{equation}
where $E_{j}$ are again the components of the\emph{ }incident field,
and coefficients $\alpha_{ijk}$, related to the components of the
third-rank piezoelectric tensor \cite{Fridkin2001}, are non-zero
only for noncentrosymmetric media. Examples include piezoelectrics
or ferroelectrics such as $\mathrm{LiNbO_{3}:Fe}$, quartz with $F$-centers,
or $p$-GaAs \cite{Fridkin2001}. We see that the photocurrent $\mathbf{j}$
in Eq.~\eqref{eq:photogalvanic} is nonlinear (quadratic) with respect
to the field strength\textbf{ $E$}.\textbf{}

We can also rewrite Eq.~\eqref{eq:photogalvanic} in a modified form
\cite{bulkFridkinBatirov}, \begin{equation}
j_{i}=\left(\beta_{ijk}/2\right)(e_{j}e_{k}^{*}+e_{j}^{*}e_{k})S_{0},\label{eq:photogalv_intens}\end{equation}
where $e_{j}=E_{j}/|E|$ is the $j^{\text{th}}$ component of the
incident light polarization vector, and $S_{0}$ is its intensity.
In SI units, the components of the tensor $\beta$ have the dimension
of inverse volts; normalized by the absorption coefficients, $\beta_{ijk}$
are related to the Glass coefficients for the high-voltage bulk photovoltaic
effect in nonlinear crystals \cite{bulkFridkinBatirov,GlassCoefs}.
One can similarly rewrite Eq.~\eqref{eq:photogalvanic_tilde}, introducing
the plasmonic equivalent of Eq.~\eqref{eq:photogalv_intens} as\begin{equation}
j_{z}=\tilde{\beta}_{zxx}S_{0}|e_{x}|^{2},\quad\tilde{\beta}_{zxx}=\rho\xi\eta_{o}e/(\hbar\omega).\label{eq:photogalv_tilde_intens}\end{equation}
We see that Eq.~\eqref{eq:photogalvanic_tilde} is formally equivalent
to Eq.~\eqref{eq:photogalvanic}, standard for describing the photogalvanic
effect in bulk media. However, one important difference has to be
noted. In bulk media, directed photoelectrons are generated throughout
the volume of the material. They have a finite lifetime since their
initial velocity decays as they move. Hence the coefficients $\alpha_{ijk}$
in Eq.~\eqref{eq:photogalvanic} are proportional to that lifetime
\cite{bookSturmanFridkin}. In contrast, the geometry in Fig.~\ref{FIG:Schematics}b
considered here deals with the injection of directed photoelectrons
from a nanoparticle array into the surrounding medium. Hence there
is no lifetime in Eqs.~\eqref{eq:photocurrent_volume} and \eqref{eq:photogalvanic_tilde},
which is emphasized by using a tilde over the coefficients $\tilde{\alpha}_{ijk}$
in that formula. The behavior of photoelectrons after they have been
emitted is a subject for further discussion. It is expected that when
many nanoparticle layers are stacked together, the photoelectrons
can be recaptured and re-emitted, and in the case of bulk metamaterial%
\begin{comment}
during their subsequent motion, the {}``bulk-like'' behaviour would
resume, and%
\end{comment}
{} the effective lifetime for the directionally emitted electrons could
be reintroduced.

\section{Directional photoemission \protect \\
from conical nanoparticles\label{sec:NUMERICAL}}

To confirm the predicted photogalvanic effect in a plasmonic metamaterial,
we consider an array of gold nanoparticles whose shape is gradually
varied from cylindrical to conical (Fig.~\ref{FIG:Schematics}b--c).
The choice of conical nanoparticles is well-motivated from a fabricational
standpoint, since nanodisks fabricated using lithographic means often
acquire asymmetry and resemble cone-like shapes \cite{pracAtwaterAM10,revAtwater};
other asymmetric shapes such as hemispheres and nanopyramids are also
readily fabricable using various techniques \cite{as1-Nakayama,as2-Opal,as3-Odom}. 

Let the nanocones under study have height $h$; the larger (bottom)
facet has radius $r$, and the radius of the smaller (top) facet is
given by $r(1-\zeta)$ so that $\zeta$ can be understood as the {}``asymmetry''
or {}``conicity'' parameter. The case $\zeta=0$ corresponds to
the centrosymmetric, disk-shaped particles, whereas the opposite case
$\zeta=1$ corresponds to maximally asymmetric nanocones. For the
computational example, we use $r=25$~nm and $h=18$~nm. 

\begin{figure}
\includegraphics[width=1\columnwidth]{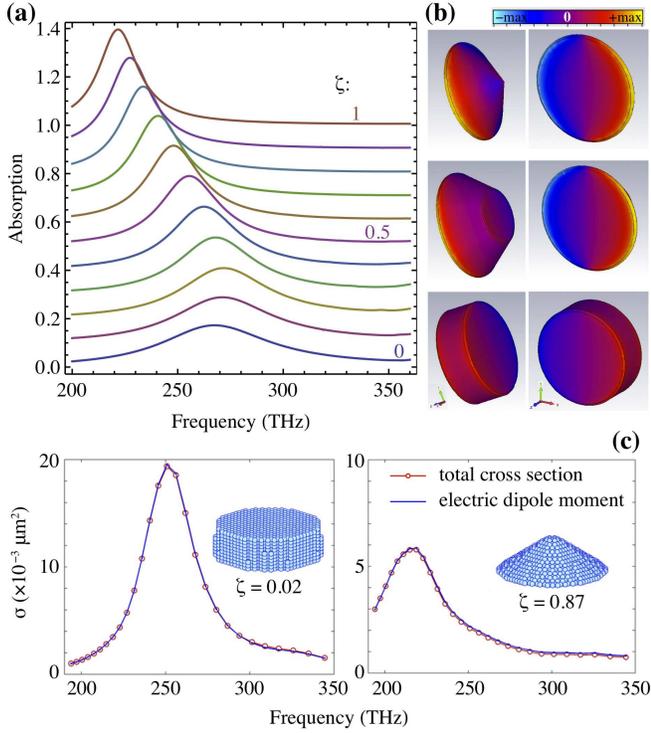}

\caption{(a) Absorption spectra for the nanoparticle lattice with the conicity
parameter $\zeta$ ranging from 0 (cylinders) to 1 (cones). The light
is normally incident onto the lattice plane and linearly polarized
along the $x$-axis. For easier readability, each line is offset by
0.1. (b) Images of the normal component of the electric field at the
plasmonic resonance (the frequency of the absorption maximum) for
$\zeta=0$, 0.5, and 1, viewed from two different angles.\emph{ }(c)
Extinction cross-section for a single nanoparticle with near-cylindrical
($\zeta=0.02$) and highly conical ($\zeta=0.87$) shape, calculated
with the DDA method comparing the full cross-section (dots) with the
electric dipole moment contribution (blue line); the insets show the
images of the discretized nanoparticles.\label{FIG:Absorption}}
\end{figure}

Permittivity of gold was described by the Drude model with plasma
frequency $2.18\times10^{15}\,\mathrm{s}^{-1}$ and collision frequency
$6.47\times10^{12}\,\mathrm{s}^{-1}$ \cite{drude}. The particles
are embedded in a homogeneous GaAs matrix ($n_{m}=3.6$, $W_{b}=0.8$
eV, $E_{g}=1.43$ eV), which results in the operating range between
870 and 1550 nm according to Eq.~\eqref{eq:energy}. For Au/GaAs
interface, the values of $\eta_{o}$ range from zero at $\hbar\omega=W_{b}$
to about 0.0025 for $\hbar\omega\lesssim E_{g}$ \cite{ourTammBulk}.
The particles are arranged in a 2D square lattice with period $a=100$~nm.
According to the earlier results \cite{plas8-Novitsky12,ourPlas},
such a dense lattice prevents the appearance of higher-order diffraction
and ensures that the lattice effects are outside of the operating
range, so the resonant mode of a particle in a lattice largely coincides
with that of an isolated particle \cite{ourPlas}. Another benefit
of making the lattice dense is the increase of the nanoparticles concentration,
resulting in the increase of the total induced photocurrent as per
Eq.~\eqref{eq:photocurrent_volume}. 

Simulations were carried out in the frequency domain using CST Microwave
Studio. The results show\textbf{ }that all the structures in question
feature a rather broad absorption resonance corresponding to the excitation
of a localized surface plasmon (Fig.~\ref{FIG:Absorption}a). As
expected, small $a$ makes the lattice-related resonances occur outside
of the operating frequency range, so the resonance is dominated by
the response of a single plasmonic nanoparticle, broadened due to
the presence of many of them \cite{ourPlas}. Note that to reduce
the influence of the discretization ({}``staircasing'') artifacts
at sharp edges (which would anyway be unphysical from the fabricational
point of view), all edges of the nanoparticles were smoothed with
the curvature radius of $\delta r=1$~nm. Together with the adaptive
mesh refinement, this proved successful in eliminating numerical artifacts
from meshing-related {}``hot spots''. 

We further see that the resonance strongly depends on the nanoparticle
shape \cite{plas8-Novitsky12}. As the shape changes from a cylinder
to a cone, the resonance undergoes a very slight blue shift up to
$\zeta=0.4$; further increase of $\zeta$ leads to a strong red shift
accompanied by a rather significant narrowing of the resonance. Plotting
the normal component of the resonant mode field on the nanoparticle
surface (Fig.~\ref{FIG:Absorption}b), we notice that although the
field at the smaller base (or tip) of the cone becomes progressively
weaker as $\zeta$ increases, the mode maintains the characteristic
outside field pattern of a fundamental dipole resonance for all nanoparticles.
To reconfirm this, we have calculated the extinction cross-section
of conical nanoparticles using the discrete dipole approximation (DDA)
method \cite{DDA1,DDA2}. Figure~\ref{FIG:Absorption}c shows that
extinction properties of the nanoparticles are fully reproduced if
only the electric dipole moment is kept in the cases of both low and
high asymmetry. 

The DDA results reproduce the red shift of the resonance as $\zeta$
increases, which is qualitatively similar to what happens in a metal
spheroid as it becomes more oblate \cite{DDAspheroid}. Moreover,
the results in Fig.~\ref{FIG:Absorption}c show the decrease of the
effective dipole moment of a nanocone compared to a nanodisk. This
decrease diminishes the coupling between the individual particles
in the lattice, weakening the absorption peak broadening for larger
$\zeta$, which is though to be the primary mechanism of the resonance
narrowing for the nanocones.%
{}

\begin{figure}[b]
\includegraphics[bb=10bp 0bp 465bp 247bp,width=1\columnwidth]{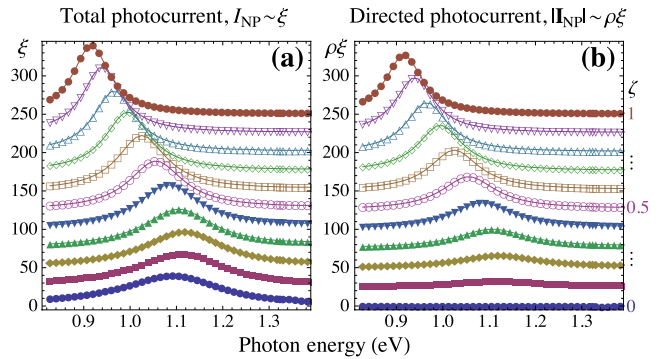}

\caption{(a) Total photocurrent $I_{NP}$ and (b) $z$-component of the directional
photocurrent $\mathbf{I}_{NP}$ for different values of the conicity
parameter $\zeta$. The lines are offset by 25 for easier readability.\label{FIG:Emission}}
\end{figure}

From the distribution of the local field at the nanoparticle surface,
which was determined numerically at every frequency, we then calculate
the photoemission current from each nanoparticle using Eqs.~\eqref{eq:photocurrent_scalar}
and \eqref{eq:photocurrent_vector}. The results are shown in Figs.~\ref{FIG:Emission}a--b,
respectively. It can be seen that the cylindrical nanoparticles have
no preferred direction of the emitted photoelectrons. However, the
more the nanoparticle shape evolves towards conical with $z$-axis
symmetry, the greater directionality along the $z$-axis the photoelectrons
acquire. 

\begin{figure}[b]
\includegraphics[width=1\columnwidth]{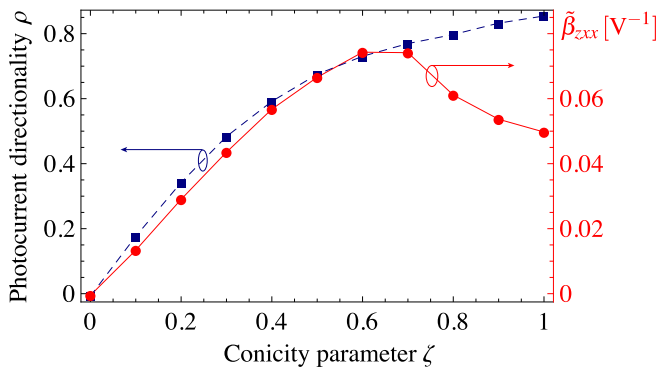} 

\caption{Photocurrent directionality ratio $\bar{\rho}$ averaged for the photon
energies between 1.0 and 1.1 eV (squares) and calculated value of
the effective $\tilde{\beta}_{zxx}$ for the value of $\eta_{o}$
for the photon energy of maximum photocurrent (the peak in Fig.~\ref{FIG:Emission}a)
calculated according to Ref.~\cite{ourTammBulk} (circles), depending
on the conicity parameter $\zeta$.\textbf{ }\label{FIG:Ratio}}
\end{figure}

The sign of $\mathbf{j}_{z}$ indicates that electrons tend to be
emitted in the direction of the base of the cone. This can be explained
by the field distributions in Fig.~\ref{FIG:Absorption}b. In a nanodisk,
the two facets have equal field distribution and therefore photocurrent
resulting from emission from the two facets balances each other; the
same happens with the side walls, resulting in overall $\mathbf{I}_{NP}=0$.
In a cone, the field distribution (and hence, photoemission) at its
base is similar to the facet of the nanodisk, but the field at the
remaining surface of the cone is significantly weaker compared to
the base because the center of mass of a nanocone gets shifted towards
its base as its asymmetry increases, again qualitatively similar to
an oblate metal spheroid \cite{DDAspheroid}. Therefore, as $\zeta$
increases, the smaller facet of the cone has a gradually declining
contribution to the total photoemission process, thus increasing the
photocurrent directionality. 

Since the operating frequency range only contains one plasmonic rsonance,
the increase in $\rho$ is almost uniform across the spectrum. We
can thus introduce an overall spectrally averaged $\bar{\rho}$ for
the structure with a given $\zeta;$ the resulting dependence $\bar{\rho}(\zeta)$
is shown in Fig.~\ref{FIG:Ratio}. We can see that highly asymmetric
nanoparticles ($\zeta>0.5$) display significant $\rho$ that exceeds
0.5, while for fully conical nanoparticles $\rho>0.8$. 

Moreover, we see that the field enhancement $\xi$ also becomes greater
when particles become more asymmetric, with maximum $\xi$ changing
from about 40 for cylinders to almost 100 for cones. Using Eq.~\eqref{eq:photogalv_tilde_intens}
and the expression for $\eta_{o}$ for Au/GaAs interface calculated
in \cite{ourTammBulk} at the maximum of the dependence $\xi(\hbar\omega)$
(see Fig.~\ref{FIG:Emission}a), we can finally derive the effective
tensor component $\tilde{\beta}_{zxx}$, also shown in Fig.~ \ref{FIG:Ratio}.
We see that the maximum value of around 0.07 is reached for conicity
parameter values $\zeta\simeq0.6\ldots0.7$ (Fig.~\ref{FIG:Ratio}).
For more asymmetric shapes, the increase in $\xi$ and $\rho$ is
compensated by the shift of plasmonic resonance towards $\hbar\omega=W_{b}$,
where $\eta_{o}$ rapidly approaches zero, leading to a slight decrease
in $\tilde{\beta}_{zxx}$ to values around 0.05. Changing the nanoparticle's
aspect ratio in such a way that its plasmonic resonance gets shifted
to higher frequencies (exceeding 1 eV) is expected to counteract this
effect and further boost the directional photocurrent. 

The resulting values of $\tilde{\beta}_{zxx}$ in Fig.~\ref{FIG:Ratio}
are seen to greatly exceed the typical values for ferroelectric crystals
such as the experimentally determined $\beta_{xxx}=3.1\times10^{-12}\,\mathrm{V}^{-1}$
in $\textrm{La}_{3}\textrm{Ga}_{5}\textrm{Si}\textrm{O}_{14}:\textrm{Fe}$
\cite{bulkFridkinBatirov}. The obtained $\tilde{\beta}_{zxx}$ is
also found to exceed the anomalously high values for bismuth ferrite
known to outperform typical ferroelectric materials by about five
orders of magnitude in the thin-film configuration ($\beta_{ijj}$
around $2\ldots3\times10^{-4}\,\mathrm{V}^{-1}$ according to the
recent measurements \cite{bulkBiFe} and first-principle theoretical
calculations \cite{bulkYoungPRL}). Furthermore, it exceeds the values
$\beta_{ijj}\sim10^{-4}\ldots10^{-3}\,\mathrm{V}^{-1}$ calculated
for the photogalvanic effect based on ratchet photocurrent from interaction
of free electrons with asymmetric nanoscatterers without taking any
plasmonic effects into consideration \cite{galvPhysE}.

Hence, the plasmonic photogalvanic effect in metamaterials based on
asymmetrically shaped nanoparticles can be confirmed to constitute
a {}``giant'' version of bulk photovoltaic effect present in non-centrosymmetric
crystals.

\section{Plasmonic versus conventional photogalvanic effect\label{sec:DISCUSSION}}

Having established formal equivalence between the conventional bulk
photovoltaic effect in non-centrosymmetric crystalline materials \cite{Fridkin2001}
and the reported plasmonic effect in nanoparticles through the parallels
between Eqs.~\eqref{eq:photogalvanic} and \eqref{eq:photogalvanic_tilde},
or similarly between Eqs.~\eqref{eq:photogalv_intens} and \eqref{eq:photogalv_tilde_intens},
we would like to discuss the similarities in the underlying physics
behind these two effects in more detail. 

Conventional \emph{photogalvanic effect} \cite{galvUFN} or \emph{bulk
photovoltaic effect} \cite{Fridkin2001} (sometimes called high-voltage
bulk photovoltaic effect \cite{bulkGlass}) refers to the generation
of intrinsic photocurrents occurring in single-phase materials without
inversion symmetry \cite{bulkYoungPRL1}. Microscopically, it is associated
with violation of the principle of detailed balance for photoexcited
non-equilibrium carriers in noncentrosymmetric crystals: the probability
of the electron transition between the states with momentum $k$ and
$k'$, $W(k,k')$, does not, in general, equal the probability of
the reverse transition: $W(k,k')\ne W(k,-k')$ \cite{galvIOP}. This
gives rise to a flux of photoexcited carriers, which manifests itself
as photocurrent with a certain direction even though the medium is
homogeneous and uniformly illuminated. 

The principle of detailed balance may be violated due to a variety
of mechanisms, e.g., inelastic scattering of carriers on non-centrosymmetric
centers, excitation of impurity centers with an asymmetric potential,
or hopping mechanism that acts between asymmetrically distributed
centers \cite{Fridkin2001}. Other effects that have been pointed
out are excitation of non-thermalized electrons having asymmetric
momentum distribution due to crystal asymmetry, delocalized optical
transitions in lattice excitation of pyroelectrics \cite{bulkGlass},
spin-orbital splitting of the valence band in gyrotropic media \cite{Fridkin2001},
and second-order nonlinear optical interaction known as {}``shift
currents'', shown to be the dominating photocurrent cause in ferroelectrics
\cite{bulkYoungPRL,bulkYoungPRL1}. 

The latter effect is strikingly similar to the plasmonic effect reported
here because the expression for the {}``shift current'', $J_{q}=\sigma_{rsq}E_{r}E_{s}$,
is essentially coincident with Eqs.~\eqref{eq:photogalvanic_tilde}--\eqref{eq:photogalvanic};
in both cases, the photocurrent is quadratic with respect to the field
strength of incident light \cite{plas5-UskovUFN12}. 

Another striking similarity arises when one compares the considered
photoemission from nanoparticles with photoionization from atoms.
%
\begin{comment}
between the phenomenological consideration of photoexcited carrier
scattering on non-centrosymmetric centers compared to (inelastic)
photon scattering on asymmetric nanoparticles.%
\end{comment}
{}Indeed, one can regard nanoparticles as atoms whose electrons are
placed into a highly asymmetric potential well, which is formed by
the boundary of metal with the surrounding medium. Then, it is known
from atomic physics that the pattern of photoeffect from such atoms
would depend on the shape of the atomic potential well. From this
point of view, it is clear that changing the shape of the nanoparticles
can efficiently deform the pattern of electron photoemission from
such nanoparticles in much the same way as what happens in non-centrosymmetric
crystals where asymmetry is inherent in the crystal lattice structure.

Hence, we have solid grounds to regard the reported plasmonic effect
in nanoparticle arrangements as the plasmonic analogy (or {}``metamaterial
counterpart'') to the conventional photogalvanic or bulk photovoltaic
effect, if we think of nanoparticles as {}``meta-atoms'' and equate
the absence of the center of symmetry in them to a similar geometric
feature of crystal lattice in bulk media. In some ways, it resembles
the \emph{mesoscopic photovoltaic effect} that was reported earlier
to occur in ensembles of semiconductor microjunctions of larger dimensions
(about $1\,\mu\text{m}$) in the microwave range \cite{meso1JETP,meso2PRB}.

Choosing between the terms \emph{photogalvanic} and \emph{bulk photovoltaic}
to name the reported effect is worth another discussion. In homogeneous
media, these two terms are essentially synonymous and have been used
interchangeably. Indeed, historically the words {}``galvanic'' and
{}``voltaic'', attributed to Luigi Galvani and Alessandro Volta
respectively (both of whom were behind the invention of a battery),
have the same meaning. In later use, though, the term \emph{photovoltaic
effect} gained a much wider recognition and at the same time, became
much more generic; it came to mean \emph{any} effect of electric energy
generation as a result of light illumination (perhaps with the exception
of photoelectron emission into vacuum, for which the term \emph{photoelectric
effect} remains more popular). Still, the predominant usage realm
of photovoltaic effects became that of the processes in the modern-day
semiconductor solar cells, i.e., the effects related to generation
and subsequent separation of electrons and holes in semiconductor
structures. To distinguish these heterostructure effects from photocurrent
generation \emph{in the bulk} of a non-centrosymmetric homogeneous
medium, the latter adopted the name {}``\emph{bulk} photovoltaic
effect''; its much less popular synonym {}``photogalvanic effect''
has not needed this addition.

For this reason, in our attempt to classify the predicted plasmonic
effect, adopting the name {}``plasmonic bulk photovoltaic effect''
can be confusing because one is tempted to forget that a plasmonic
metamaterial is only {}``effectively bulk'' in the sense that a
macroscopic excitation such as an incident plane wave will not discriminate
the individual nanoparticles and will interact with the metamaterial
\emph{as if it were} homogeneous. Microscopically, though, it is not
homogeneous; the very existence of localized \emph{surface} plasmon
excitations imply that there are surfaces that give rise to them.
The wording {}``bulk plasmonic photovoltaic effect'' would be even
more dangerous because it may mislead one into thinking that bulk
plasmons, rather than surface plasmons, are at work, which is not
the case. Therefore, to avoid such confusions, we argue that \emph{plasmonic
photogalvanic effect} is the proper name for the reported phenomenon. 

In the broader picture, it has attributes of both a bulk effect and
a surface effect depending on the length scale, and in this way it
bridges the gap between the inner photoelectric effect (defined as
electric charge carrier generation due to photon absorption in a bulk
material) and the outer photoelectric effect (defined as electric
charge carrier emission from one medium into another across an interface)
\cite{bookInnerOuter}; this is not to be confused with internal vs.~external
photoelectric effect, both of which are subclasses of the outer photoelectric
effect depending on whether the second medium is solid or not \cite{bookInternalExternal}.

One does need to keep in mind, however, that the present numerical
demonstration of plasmonic photogalvanic effect, based on a 2D arrangement
of nanoparticles, has only succeeded in demonstrating the equivalence
for a {}``thin slab'' of metamaterial, analogous to a thin film
of a bulk non-centrosymmetric crystalline medium. Further comparison
of plasmonic and conventional photogalvanic effect involving a 3D
arrangement of nanoparticles should therefore be forthcoming. It is
rather straightforward, the key challenge being the means to provide
uniform illumination in a medium with sufficiently many lossy metallic
inclusions under the condition of localized plasmonic resonance (and
hence, highly inhomogeneous and strongly enhanced local fields). Overcoming
this challenge may result in a cap on the maximum nanoparticle density
and therefore, in a limit on how strong plasmonic photogalvanic effect
can be.

That said, the results presented in the numerical demonstration in
Section~\ref{sec:NUMERICAL} point out that in terms of the relevant
tensor components, plasmonic photogalvanic effect can be several orders
of magnitude stronger than the conventional one. Hence, calling the
effect \emph{giant plasmonic photogalvanic effect} is warranted, on
par with giant magnetooptical effects present in chiral metamaterials
and similarly surpassing the naturally occurring chirality in bulk
media by orders of magnitude \cite{GiantGyro}.

\section{Conclusions and outlook\label{sec:CONCLUSIONS}}

To summarize, we have theoretically predicted new functionality in
photoconductive metamaterials: the \emph{giant plasmonic photogalvanic
effect}, analogous to photogalvanic (or bulk photovoltaic) effect
in homogeneous non-centrosymmetric media \cite{bookSturmanFridkin}.
The reported effect is numerically demonstrated in a metamaterial
containing similarly oriented non-centrosymmetric metallic nanoparticles
embedded in a homogeneous semiconductor matrix, when illuminated by
a wave with photon energies insufficient for the internal photoelectric
excitation in the semiconductor. Due to the lower degree of symmetry
in the nanoparticles (the absence of mirror symmetry in the $x$-$y$
plane), the flux of {}``hot'' electrons emitted from the nanoparticles
with the assistance of a resonant plasmonic excitation aquires directionality.
Averaged over the volume of the metamaterial, this directionality
is manifest as an electromotive force resulting from homogeneous external
light illumination (the photogalvanic effect). 

We have also found that the resulting current density generated in
a layer of nanoparticles grows as the particle shape changes from
cylindrical to conical%
\begin{comment}
, i.e., with the increase of the particle asymmetry as parametrized
by $\zeta$ in Fig.~\ref{FIG:Schematics}%
\end{comment}
{}. Furthermore, we have calculated the component $\tilde{\beta}_{zxx}$
of the effective third-rank tensor that relates the induced current
density to the incident electric field intensity {[}see Eq.~\eqref{eq:photogalv_tilde_intens}{]}.
We have shown that the effective $\tilde{\beta}_{zxx}$ for the nanoparticle
array exceeds the components $\beta_{ijk}$ for the naturally occurring
ferroelectrics with bulk photovoltaic effect \cite{bulkBiFe,bulkYoungPRL,bulkFridkinBatirov}
by orders of magnitude. Hence, the reported plasmonic effect can be
regarded as a {}``giant'' version of the photogalvanic effect occurring
in natural materials, adding to the assortment of effects that are
much stronger in artificial metamaterials than in natural media. %
\begin{comment}
Note that nanoparticles can be treated either within the metamaterial
framework, i.e., as quasi-microscopic objects interacting with light
as a whole, or as individual macroscopic objects. Dependent on the
treatment chosen, the reported plasmonic photogalvanic effect can
be classified as either the inner photoelectric effect, which occurs
as a result of light-matter interaction on the level of a single microscopic
object (e.g., atom, molecule or impurity center in semiconductor),
and the outer photoelectric effect, which occurs at material interfaces
in a macroscopic setting \cite{bookInnerOuter}. Hence, the reported
effect serves to bridge the gap between the inner and outer photoelectric
effect, helping us to better understand the nature of either one on
a fundamental level and potentially providing important clues regarding
the nature of plasmon-assisted electron photoemission.%
\end{comment}
{}

On a fundamental level, the proposed effect is important for our understanding
of plasmon-assisted electron photoemission processes, and constitutes
a new way of exploring light-matter interaction at the mesoscopic
scale. On a more applied level, our results can be used in a variety
of ways, from a new way of characterizing plasmonic structures (distinct
from purely optical or electron-microscopy approaches), to new designs
of photodetectors operating outside of the spectral range for band-to-band
transitions for semiconductors. It can also be used to increase the
performance of photovoltaic elements by making use of longer-wavelength
photons, which are normally lost in traditional cells based on the
inner photoelectric effect \cite{uHalasSci11,plas8-Novitsky12}. The
result that photoemission predominantly occurs at the base of the
nanocones makes them particularly appealing for photovoltaic devices
where nanoparticles are deposited on a semiconductor substrate. 

Moreover, we can regard photoemission in non-centrosymmetric plasmonic
nanoantennas as a {}``ratchet'' (Brownian-motor) mechanism \cite{ratchetReimann,ratchetHanggi}
that works as an optical rectifier or {}``rectenna'' \cite{ratchetGreen}.
The fundamental concept of optical ratchet devices and optical rectennas
attracts much attention in recent developments in nanotechnology \cite{ratchetAppErman,ratchetAppSassine,ratchetAppValev,ratchetAppHatano}. 

Finally, we note that we have only considered the plasmonic analogue
of the linear bulk photovoltaic effect, since the nanoparticle shape
was chosen to be achiral. It is expected that chiral (or planar chiral)
nanoparticles would provide the plasmonic analogue to the circular
bulk photovoltaic effect \cite{bookSturmanFridkin,bulkFridkinBatirov},
described on analogy with Eq.~\eqref{eq:photogalv_intens} as \begin{equation}
j_{i}^{C}=i\beta_{ij}^{C}\left[\mathbf{e}\times\mathbf{e}^{*}\right]_{j}S_{0}.\label{eq:photogalv_cirular}\end{equation}
Thus, designing plasmonic nanostructures with anomalously high effective
$\tilde{\beta}_{ij}^{C}$ is expected to result in new ways to characterize
both chiral plasmonic nanostructures and chirality-related properties
of light. 
\begin{acknowledgments}
The authors would like to thank Jesper Mørk for valuable comments.
S.V.Z. acknowledges support from the People Programme (Marie Curie
Actions) of the European Union\textquoteright{}s 7th Framework Programme
FP7-PEOPLE-2011-IIF under REA grant agreement No.~302009 (Project
HyPHONE). V.E.B. acknowledges support from SPIE Optics and Photonics
Education Scholarship, as well as Otto Mønsteds and Kaj og Hermilla
Ostenfeld foundations. I.E.P. and A.V.U. acknowledge support from
the Russian MSE State Contract N14.527.11.0002 and from the CASE project
(Denmark).\end{acknowledgments}

\end{document}